\documentclass[oneside,onecolumn]{aa}
\usepackage{times}
\usepackage[T1]{fontenc}
\usepackage[latin1]{inputenc}
\usepackage{graphics}
\usepackage{psfig1}

\makeatletter

\def\be{\begin{equation}}
\def\ee{\end{equation}}
\newcommand{\varthetavec}{{\mbox{\boldmath$\vartheta$}}}

\newcommand{\ave}[1]{\left\langle #1 \right\rangle}

\makeatother

\begin{document}

\title{A conspicuous tangential alignment of galaxies in a STIS
Parallel Shear Survey field: A new dark-lens candidate ? \thanks{Based on
observations made with the NASA/ESA \emph{Hubble Space Telescope}, obtained at
the Space Telescope Space Institute (STScI) which is operated by AURA Inc.,
under NASA contract NAS 5-26555}}

\author{J.-M. Miralles\inst{1,2}, T. Erben\inst{1,3,4}, H.
H\"ammerle\inst{1,6}, P. Schneider\inst{1}, R.A.E. Fosbury\inst{2}, W.
Freudling\inst{2}, N. Pirzkal\inst{2}, B. Jain\inst{5} \and S.D.M.
White\inst{6}.}

\institute{Institut f\"ur Astrophysik und Extraterrestrische
Forschung, Universit\"at Bonn, Auf dem H\"ugel 71, Bonn, Germany \and ST-ECF,
European Southern Observatory, Karl Schwarzschild Str. 2, Garching b.
M\"unchen, Germany \and Institut
d'Astrophysique de Paris, 98bis Boulevard Arago, Paris, France \and
Observatoire de Paris, DEMIRM 61, Avenue de l'Observatoire, Paris, France \and
University of Pennsylvania, Dept. of Physics and Astronomy, Philadelphia, USA
\and Max-Planck Institut f\"ur Astrophysik, Karl Schwarzschild Str. 1,
Postfach 1317, Garching b. M\"unchen, Germany}

\offprints{J.-M. Miralles \email{miralles@astro.uni-bonn.de}}


\date{Received / Accepted}

\abstract{
We report the serendipituous discovery of a conspicuous alignment of galaxies
in a field obtained through the STIS Parallel Shear Survey. This project
collects randomly distributed $50''\times 50''$ fields to investigate the
cosmic shear effect on this scale. Analyzing the parallel observations having
the Seyfert galaxy NGC625 as primary target, we recognized over the whole field
of view a strong apparent tangential alignment of galaxy ellipticities towards
the image center. The field shows several arclet-like features typical for
images of massive galaxy clusters, but no obvious over-density of bright
foreground galaxies. We also find a multiple image candidate.
On the basis of the possible strong and weak lensing effect within the data,
we discuss whether this could be compatible with a massive halo with no
clear optical counterpart.  }

\authorrunning{J.-M. Miralles et al.}
\titlerunning{A conpicuous tangential alignment of galaxies in a SPSS field:
A new dark lens candidate ?} 
\maketitle

\section{Introduction}

The STIS CCD camera on-board Hubble Space Telescope has been providing
an unique view of faint galaxies. The characteristics of this
instrument which offers good depth, excellent resolution and a good
sampling of the HST PSF, together with the possibility of obtaining
data using the parallel observing mode, makes it a suitable instrument
for studies such as the STIS Parallel Shear Survey (hereafter refered
as SPSS), whose main goal is the study of cosmic shear on scales below
one arcminute. Data for this program started to be collected on
September 24th, 2000, within the scope of an HST GO cycle 9 parallel
proposal (8562 and 9248, P.I.: P. Schneider) and up to date, more than
500 different pointings have been obtained. The use of STIS CCD
imaging in the CLEAR filter mode allows to combine good spatial
resolution (originally $0.05''$ per pixel) with very deep exposures
since the CCD is un-filtered and sensitive to wavelengths between 2500
and 11000 \AA  (Pirzkal et al. 2001). We report in this paper the
discovery of a conspicuous alignment of galaxies in a deep exposure
obtained with that program. This field contains several
high-ellipticity galaxies which seem oriented around a point near the
center of the field giving a striking visual impression that caught
our attention. Also, there are several fainter galaxies in this field
which also contribute to the visual impression of an alignment of
galaxies. We used the aperture mass technique and an analysis of the
most elliptical objects in the field to try to quantify the
significance of this effect and test the visual impression. Section
2 describes our data and the reduction procedure. The core of section
3 presents the aperture mass analysis, discussing the significance of
the peak of mass obtained and the errors on its position.  Section 4
presents the most elongated objects in the field, and discusses their
orientation with respect to the peak found with the aperture mass
technique. We discuss also in this part the similar characteristics of
2 of these candidate arclets that could be compatible with being
multiply imaged. Finally, section 5 summarizes and discusses the
indications presented in the paper which, taken together, could make a
case for the presence of a possible underluminous lens in this field.

\section{Data}

These data were obtained as part of the large HST GO cycle 9 parallel
proposal aimed at the study of cosmic shear (see Pirzkal et al. 2001
and http://www.stecf.org/projects/shear for more details on the STIS
data characteristics). The field considered here was, coincidentally,
the first one observed as part of that program. The associated
image is composed of 18 individual 400s exposures with a field of view
of $50'' \times 50''$, taken in the CLEAR filter mode. The exposures were
individually cleaned for hot pixels and cosmic rays using the Eye
algorithm in SExtractor (Bertin \& Arnouts 1996)\footnote{Eye and
SExtractor are part of the Terapix software suite and can be found at
http://terapix.iap.fr/soft/}, which creates masks of the pixels that
are not used during the combination, eliminating 99\% of the
spurious pixels. The shifts between the different exposures were
computed using the method described in Pirzkal et al.\ (2001), which
can align these images to an accuracy of 1/10 of a
pixel. Finally, the individual exposures were drizzled to a pixel
size of $0.025''$ (Fruchter and Hook 2002) and combined by median
averaging with the IRAF Imcombine procedure, achieving a total
exposure time of 7200s. The use of this reduction procedure, basically
similar to the one described in Pirzkal et al. (2001), is known to
preserve the original PSF shape and size (H\"ammerle et al. 2001)
which is extremely important for an accurate analysis of the
shear. The coordinates of the field are R.A: $01^{\rm h} 35' 06.7''$
and Dec: $-41^o 21' 20.18''$ (J2000), about $7'$ north of the local
Seyfert galaxy NGC625, which was the primary target observed with the
WFPC2. Throughout this paper we will refer to this field as the Slens1
field.  The coadded image is available in fits format at
http://www.stecf.org/projects/shear/slens/slens.fits. The SExtractor
parameter file used is also available there in order to enable others
to reproduce the SExtractor catalog.

\section{$M_{\rm ap}$ statistical analysis}

\begin{figure}[ht]
\centerline{\psfig{figure=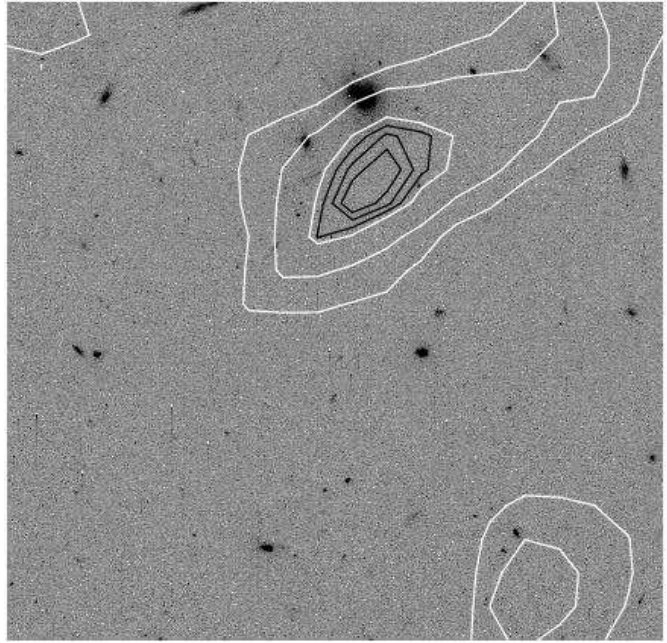,height=0.35\textheight,width=\hsize,angle=-90}}  
\caption{The figure shows contour maps of the $S/N$ of $M_{\rm ap}$ values
(eq. \ref{snestimate}) superimposed on the field. The $S/N$ ratio was
calculated on a 20x20 grid of our image. The filter scale $\theta$ is 
$25''$. White contours correspond to $S/N=1.0, 1.5, 2.0$ and black
contours to $S/N=2.1, 2.2, 2.3$. 
See the text for a discussion on the robustness of this result.}
\label{mapsnfig}
\end{figure}

\begin{figure*}[ht]

\centerline{\psfig{figure=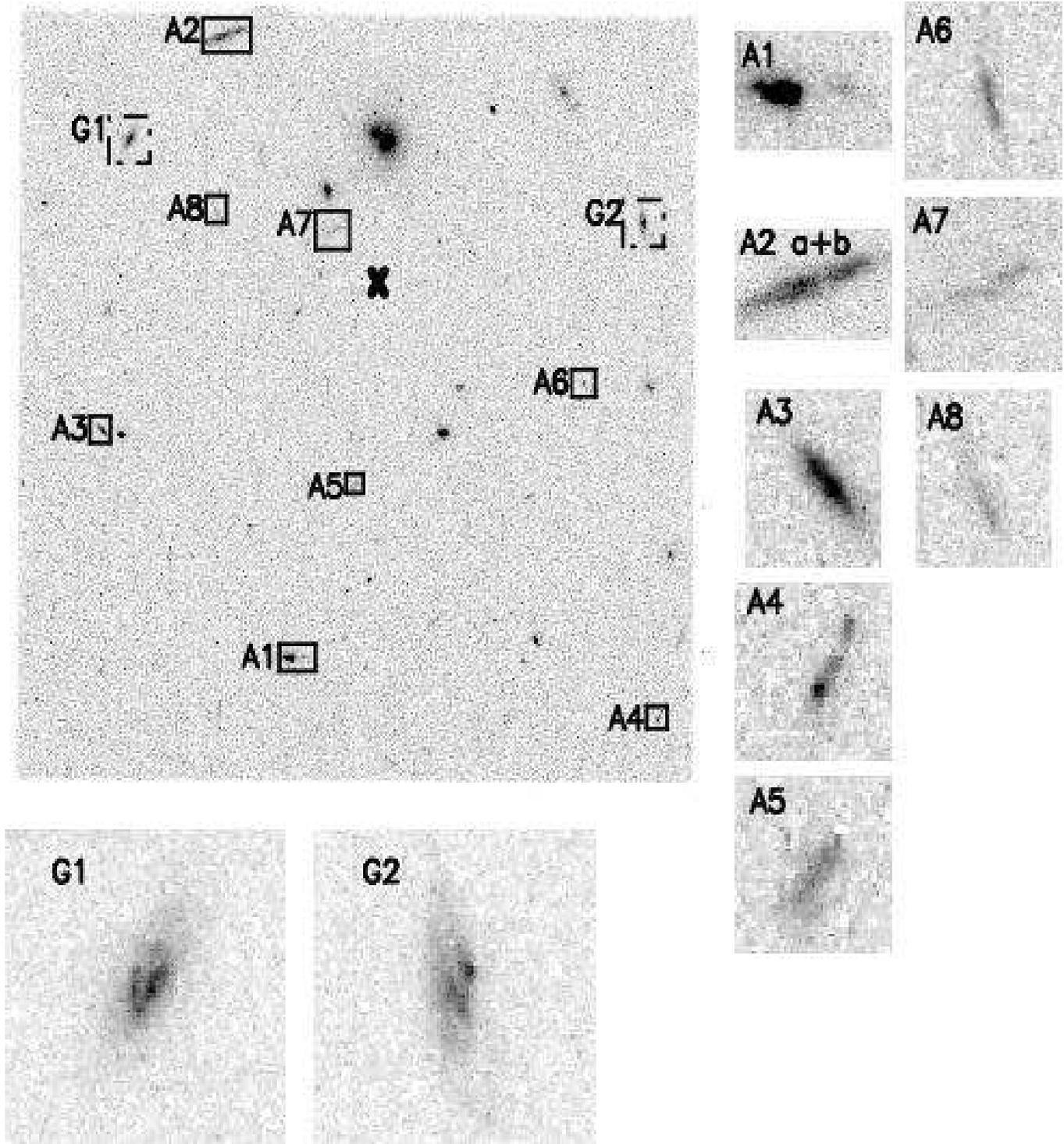,height=0.8\textheight,width=\hsize,angle=-90}} 
\caption{This figure shows the field with a zoom on the two candidates for
multiply-imaged galaxies G1 at the left and G2 at the right side of the field
and 9 other arclet-like candidates (A1-A8) with isophotal AB magnitudes 
ranging from 22.7 to 26.9. Note that the 2 components of A2 (A2a and A2b) are
considered as 2 different objects. The `X' marks the $M_{\rm ap}$ 
peak of the $25''$
smoothing of Fig. 1. The 2 multiple image candidates are located on each side
of the brightest galaxies. They show a very characteristic V-shape morphology
and have the same surface brightness.}  \label{objectsim} 
\end{figure*}

Over the whole field of view of our $50'' \times 50''$ STIS image, we 
recognize a strong apparent tangential alignment of galaxies
towards the center (see Figs. \ref{mapsnfig} and
\ref{objectsim}). Although the small $50''\times 50''$ field does not
allow a quantitative weak lensing analysis to verify or to falsify the
hypothesis of a massive dark matter halo in the image center, we use
the $M_{\rm ap}$ statistic (Schneider 1996) to test whether we can
identify a significant and stable peak.  The
details of the creation of an object catalog are the same as for the
rest of the SPSS project. They are described and justified in depth in
H\"ammerle et al. (2001). In short, we performed an
analysis of the data using two different software packages.  
Objects were detected with SExtractor and a
modified version of the IMCAT package (Kaiser, Squires \& Broadhurst
1995 hereafter referred to as KSB; Erben et al. 2001). Finally, we only
retained objects detected with both programs, assigning them the
photometric information from SExtractor and the shape parameters from
IMCAT. There are not enough foreground stars in a STIS image to allow
directly a PSF correction in the KSB style from the data itself, which is
the usual procedure with wide-field ground-based observations.  On the other
hand, our analysis of STIS data, especially starfields obtained over
the same period of time as our observations, showed that the PSF is
sufficiently stable (H\"ammerle et al. 2001) so that these starfields
provide a good estimate of the quantities used for the
KSB anisotropy and smearing correction (see also Hoekstra et al. 1998
for a similar procedure with WFPC2 data). The final correction was
performed using the PSF information from the starfields as described
in Erben et al. (2001). We removed from the catalog the objects with
an axis ratio larger than two; those will be considered in the next
section as candidate arclets. Their alignment towards the direction
defined by the $M_{\rm ap}$ statistics will provide an independent
test of the shear signal. In the end, our catalog contained 54
objects, with fully corrected ellipticities, which were used for the
subsequent analysis.

We use standard lensing notations in this paper. For a broader introduction to
this topic, the reader should refer to Bartelmann \& Schneider (2001).
We used the $M_{\rm ap}$ statistic (Schneider 1996) for the weak lensing
analysis. The family of $M_{\rm ap}$ statistics uses the fact that a filtered
integral over the convergence $\kappa$ can be converted into a filtered
integral over the observable tangential shear $\gamma_{\rm t}$ [see equations
(9) and (15) in Schneider 1996], if the filter function $U(|\varthetavec|)$
satisfies $\int_0^\theta{\rm d}\theta'\;\theta'\,U(\theta')=0$ but is arbitrary
otherwise. It is straightforward to construct
an unbiased estimate $M_{\rm ap}'$ for the integral by a discrete sum over
observed image ellipticities $\epsilon_{{\rm t}}$ and considering the coordinates origin being at the center:
\be
  M_{\rm ap}'(\theta)={\pi\theta^2\over N}\sum_i \epsilon_{\rm
  t}(\varthetavec_i)\, Q(|\varthetavec_i|)\;
  \label{mapestimatedata}
\ee
with $Q(x)=q(x/\theta)/\theta^2$ and 
$q(\rho)=(6/\pi)\rho^2(1-\rho^2)$. 

Moreover, as it is a scalar quantity, expressions for the variance and hence
the $S/N$ for a derived $M_{\rm ap}$ value are easily calculated. For
the $S/N$ one obtains (see Schneider 1996):

\be
  S={\sqrt{2}\over \sigma_\epsilon}
  {\sum_i {\epsilon_{\rm t}(\varthetavec_i)}
  \,Q\left(|\varthetavec_i|\right) \over
  \sqrt{ \sum_i {Q^2\left(|
  \varthetavec_i|\right)}}},
  \label{snestimate}
\ee
where $\sigma_{\epsilon}$ is the intrinsic dispersion of image
ellipticities.  It is estimated by the dispersion of the galaxies
entering the calculation of $S$ [$\sigma_{\epsilon}^2=N^{-1}\sum_i
(\epsilon_{{1}i}^2+\epsilon_{{2}i}^2$)].  Hence the $M_{\rm ap}$
statistics are the ideal tool to evaluate the significance and
robustness of a mass concentration.  In our case, with only a
$50''\times 50''$ field and 54 objects in hand, the possibilities for
this kind of analysis are limited. We are confined to small radii for
our filter scale $\theta$ (about $25''$) for which the $M_{\rm ap}$
statistic is extremely noisy. Moreover, applying
(\ref{mapestimatedata}) to a grid of points on our data will not give
an unbiased estimate of $M_{\rm ap}$ since the data do not cover a
complete circle with radius $\theta$ at most positions. However, even
if $M'_{\rm ap}$ can no longer be interpreted as an unbiased estimate
of the filtered density profile, it yields a measure of the tangential
alignment around a point and as such is a useful statistical quantity.

The application of this approach to our data gives the results shown
in Fig.  \ref{mapsnfig}. There, we show a significance map for one
filter radius ($\theta = 25''$). We could slightly vary the size of
the filter, but we are limited by the size of the field and the number
of objects in it.  Changing the filter size down to $20''$ or up to
$30''$ does not modify the results. We recover a peak located at the
image pixel coordinates $x=1066; y=1297$ (here and in the following,
`pixels' denote the subsampled pixels of size $0.025''$ and with an origin 
for the coordinate system at the lower left corner of the image), whose
position approximately coincides with the position of the brightest
galaxy in the field. With a $S/N\approx 2.42$ for this smoothing, the
peak is marginally significant. Here, the significance was estimated
with (\ref{snestimate}).  We found that these estimates are in very
good agreement with those obtained from randomizing the orientations
of galaxy ellipticities.  To judge whether the peak is a robust
statistical signal, we repeated the calculation of the $M_{\rm ap}$
statistics drawing randomly $70\%$ and $50\%$ of the galaxies out of
the initial 54 objects catalog. This exercise was repeated 100 times
and the positions of the most significant peaks in the realizations
were considered. The mean and standard deviation for the recovered
positions of the peak are in the $70\%$ case: $\ave{x}=1250;
\sigma_x=265; \ave{y}=1383; \sigma_y=244$ and for the $50\%$ case:
$\ave{x}=1234; \sigma_x=300; \ave{y}=1286; \sigma_y=350$. The number
of "catastrophic outliers", where the most significant peak is located
at the very border of the frame, is about $12\%$ in both cases. The
mean amplitude for the peaks is $S/N\approx 2.4$ ($70\%$ case) and
$S/N\approx 2.1$ ($50\%$ case). These values are a bit higher than
expected from simple Gaussian statistics.  Also if we choose our
samples not randomly but remove the $30\%$ and $50\%$ most elliptical
galaxies from the initial catalog, we recover peaks consistent in
position with these results (although in the $30\%$ case, a strong
peak appears at the border of the frame).

Finally we checked the probability of obtaining at least a $S/N\approx
2.42$ detection in fields with randomly oriented galaxy
ellipticities. As mentioned above, equations (\ref{mapestimatedata})
and (\ref{snestimate}) applied to our field do not provide unbiased
estimates of $M_{\rm ap}$ and its uncertainty, since our data do not
cover a complete circle of the smoothing radius at the peak
position. In order to estimate this probability, we generated 10000
$50''\times 50''$ fields with 54 objects randomly placed, with random
galaxy ellipticities drawn from the distribution $P(\epsilon_1,
\epsilon_2)\propto \exp(-|\epsilon|^2/ \sigma_{\epsilon}^2)$ with
$\sigma_{\epsilon}=0.36$ matching the $\sigma_{\epsilon}$ of our
data. We find that we can reproduce a peak of at least $S/N\approx
2.42$ with $700<x_{\rm peak}<1300; 1000<y_{\rm peak}<1600$ in $2.85\%$
of all cases. We conclude that the 54 objects detected in our frame
show a fairly robust alignment towards the upper middle part of our
image. As stated above, the lack of knowledge about the shear outside
the Slens1 field does not permit decisive conclusions about the
existence of a possible massive structure within the field.

\section{Arclet-like features and multiple object candidate.} \label{arcanal}
We selected arclet candidates solely on the basis of their
elongation. The criterion choosen was that the elongation, as defined
in SExtractor (major axis over minor axis), should be greater than
two. Eleven objects met the requirement, with isophotal magnitudes
ranging from 22.7 to 27.3 (see Table 1), which are shown in Fig. 2.
A2, which appears to have 2 components, is detected by SExtractor as
two distinct objects (A2a and A2b), each with axis ratio larger than
two. As stated in section 3, these objects were not included in the
catalog used to compute the $M_{\rm ap}$ statistic. Eight of those
galaxies are distributed tangentially to the position of the peak
center at $x=1066; y=1297$, but three of them (A5, A6 and A8) are
oriented more radially. The angle between their minor axis and the
direction towards the $M_{\rm ap}'$-peak center is on average 24
degrees. The probability that the mean of this angle for 11 objects is
smaller than the measured one form randomly oriented objects is only
$0.3\%$, indicating that those galaxies are well oriented tangentially
with respect to the peak center that was defined by the less elongated
objects. This seems difficult to explain unless we consider a
gravitational effect.

\begin{table}
\caption[]{Characteristics of the objects selected as arclet candidates.
Magnitudes are AB isophotal magnitudes for the STIS CCD clear mode filter.
Positions are given in image subpixels. The elongation is defined by
major axis over minus axis. $\theta$ is the angle in degrees between the major
axis and the X-axis.}          
\label{arctab}      
$$          
\begin{array}{lcrrcr}         
   \hline             
\noalign{\smallskip}
            Name  &  Magnitude & X & Y & Elongation & \theta \\
            \noalign{\smallskip}
            \hline
            \noalign{\smallskip}

            G1 & 23.44 & 366 & 1697 & 2.1 & 61 \\
            G2 & 23.99 & 1870 & 1475 & 2.5 & -77 \\
            A1 & 22.74 & 835 & 354 & 2.0  &  -10 \\
            A2a & 23.25 & 634 & 1961 & 3.5  &  19 \\
            A2b & 24.12 & 679 & 1975 & 2.6  &  17 \\
            A3 & 24.24 & 287 & 942 & 2.7  &  -46  \\
            A4 & 25.76 & 1914 & 196 & 2.4  &   58 \\
            A5 & 25.86 & 1025 & 801 & 2.5 &  56  \\
            A6 & 26.11 & 1697 & 1062 & 3.0  & -75   \\
            A7 & 26.79 & 947 & 1454 & 2.6  &  10  \\
            A8 & 26.91 & 621 & 1503 & 3.6 &  -62  \\
            \noalign{\smallskip}
            \hline
         \end{array}
     $$
\end{table}

The high-resolution of STIS CCD imaging permits the identification of
candidate multiply imaged galaxies, even in the absence of any color
information as it is the case here, just by using morphology
and surface brightness.  The strongest case for a multiply imaged
galaxy candidate comes from objects G1 and G2 (Fig. 2). These two
galaxies show an unusual and similar V shape morphology, superimposed
on an extended and dimmer halo.  They have also equal maximum surface
brightness, 23.29 mag/arcsec$^2$ for G1 and 23.34 mag/arcsec$^2$ for
G2. They are located at opposite sides of the $M_{\rm ap}$ peak: G1
being at $19.45''$ from the position of the peak and G2 at
$22.53''$. Also their tangential alignment relative to the direction
of the peak is nearly perfect. This symmetry, combined with the
characteristic morphology and the identical surface brightness of
these objects, may indicate that they are multiple images of the same
background galaxy. However, the two images seem to show the same
parity, with a brighter star-forming region located on the right side,
when we would expect an odd parity in the case of a double image.

Without color information, the question of whether G1 and G2 are
indeed a double image is impossible to decide. The indication of the
same parity could be false if the bright spot on the right in G2 is an
unrelated object. However, if they have indeed the same parity, a lens
interpretation is difficult. Simple three-image lens models would
predict a third image within the field-of-view, unless the putative
lens is extremely massive. One such example would be a barely
critical lens, whose caustic consists only of a lips singularity (see
Schneider, Ehlers \& Falco 1992); in this case, two equal parity
images appear on either side of the lens, with the third image near
the center. No indication for such a central image is found.

\section{Discussion}

The data presented in the previous sections indicate a
systematic alignment, and possibly a multiply-imaged object in the
field. The easiest interpretation of these
findings is the existence of a mass in the line of sight of Slens1
which could produce both strong (one candidate multiple image galaxy
but with a puzzling parity problem) and weak (a coherent tangential
alignment which can be quantified through the $M_{\rm ap}$-statistics
as a significant peak together with several candidate arclets
distributed tangentially around the same position) gravitational
lensing effects. However, with our current STIS data, it is not
possible to proceed further with the analysis and to confirm
that a massive dark halo is present in this field. The significance of
the $M_{\rm ap}$ peak is too low to discard the possibility that it is
a statistical fluctuation, even if we would not expect a better
significance given the size of the field and the number of objects in
it. We therefore considered the result of the $M_{\rm ap}$-analysis
only to define a `center' of alignment, relative to which the
orientations of the arclets were measured. The mean of these
orientations is small and is expected to occur randomly with a
probability of only $0.3\%$. The arclets show a pattern strongly
resembling those of lensing clusters.

If these image features are produced by a gravitational lens,
it does not seem to be associated with any luminous objects visible in
the deep exposure obtained with STIS. Only the location of the $M_{\rm
ap}$ peak could suggest that this mass may somehow be associated with
the group of bright elliptical galaxies on the center-top part of the
field.  If we suppose that this is the case, the mass necessary to
produce the candidate multiply imaged galaxy like G1 and G2 within a
$20''$ radius would be similar to the mass of a rich cluster of
galaxies. If the lens is really associated with those bright galaxies
located close to the peak, then the M/L ratio is at least 2 orders of
magnitude above the normal values for groups of elliptical
galaxies. In any case, if this is a lens, the mass we find is
unacounted for in terms of light for what is normally regarded as the
M/L ratio in groups or clusters of galaxies. We tested also the light
profile of the brightest elliptical in this group. This galaxy could
be at a redshift of 0.5-0.6 based on similar galaxies in terms of
magnitude and apparent size for which we have spectroscopic redshifts
from the CFRS (Lilly et al. 1995). If this galaxy had a cD profile, it
would strongly suggest the presence of an underlying cluster, but this
galaxy clearly has a deVaucouleurs profile, which makes the test
inconclusive.

In conventional models of structure formation, the most massive halos
detected by aperture mass techniques like the one described in section 3 would
be easily observed because they should have accreted gas which, after
subsequent cooling, would form stars. Nevertheless, it is possible that one
could find a number of dark lenses, where the dark matter would have
collapsed in structures massive enough to create a strong lens effect, while
prohibiting baryonic matter to settle in the dark matter halos and initiating
the star formation. So far, two other candidates have been found (Erben et
al. 2000, Umetsu \& Futamase 2000) but neither of them has been clearly
confirmed. Also, a class of X-ray Over-Luminous Elliptical Galaxies (OLEGs,
Vikhlinin et al. 1999) show similar characteristics of
a dark halo, being compatible in X-ray with the mass of a poor cluster but
presenting a very low optical luminosity. Recently White et al. (2001) have
suggested that those detections could be explained by projection
effects of several mass sheets along the line of sight. Another explanation
could be that most of the objects observed in this field are at the same
redshift and belong to the same gravitational structure, the preferred
orientation of the galaxies being then  just a product of their infall into
the gravitational potential. 

To confirm the reality and the nature of the
$M_{\rm ap}$ peak, wider deep exposures centered on the position of the peak
are necessary. At the same time, wider deep multicolor imaging would
allow us to
obtain photometric redshifts that would give an indication for the redshift
of all the objects, including G1 and G2, in the field and its surroundings. If
G1 and G2 present similar photometric redshifts then, if this is a lens, a
third image should be detected in those images unless this is a very
unexpected configuration. The combination of weak-lensing analysis and
photometric redshift information can then be used to unveil the true nature
of the objects in this field. Wittman et al. (2001) have recently used the
combination of these techniques for the discovery of a new cluster of
galaxies.  The detection and confirmation of only a single dark-lens would
already have important consequences for the standard picture of structure
formation (Trentham et al. 2001). It would also show the importance of using
gravitational lensing as a tool for surveys, selecting objects by mass instead
of by light as in traditional methods. 

\begin{acknowledgements}
We would like to thank Yannick Mellier and Ludovic Van Waerbeke for
long and helpful discussions and for first pointing out at us the
double image candidate. This work was supported by the TMR Network
``Gravitational Lensing: New Constraints on Cosmology and the
Distribution of Dark Matter'' of the EC under contract
No. ERBFMRX-CT97-0172, by the DLR grant 50 OR 0106 and the Deutsche
Forschungsgemeinschaft. 
\end{acknowledgements}


\begin{thebibliography}{20}
\bibitem{1}Bartelmann, M. \& Schneider, P., 2001, Phys. Rep., 340, 291
\bibitem{2}Bertin, E. \& Arnouts, S., 1996, A\&AS, 117, 393
\bibitem{3}Erben, T., Van Waerbeke, L., Bertin, E., et al., 2001, A\&A, 366,
717 
\bibitem{4}Erben, T., Van Waerbeke, L., Mellier, Y., et al., 2000, A\&A,
355, 23 
\bibitem{5}Fruchter , A. \& Hook, R., 2002, PASP accepted,
astro-ph/9808087 
\bibitem{6}H\"ammerle, H., Miralles, J.-M., Schneider, P., et al.
2001, A\&A accepted, astro-ph/0110210
\bibitem{7}Hoekstra, H., Franx, M., Kuijken, K., Squires, G., 1998, ApJ, 504,
636
\bibitem{8}Kaiser, N., Squires, G., Broadhurst, T., 1995,
ApJ, 449, 460
\bibitem{9}Kochanek, C.S., 1990, MNRAS, 247, 135
\bibitem{11}Lilly, S., Hammer, F., Le Fevre, O., Crampton, D., 1995, ApJ, 455,
75
\bibitem{12}Mellier, Y., 1999, ARAA, 37, 127 
\bibitem{13}Pirzkal, N., Collodel, L., Erben, T., et al., 2001, A\&A 375, 351
\bibitem{14}Schneider, P., Ehlers, J. , Falco, E.E., 1992, Gravitational
Lenses (Springer: New York) (SEF) 
\bibitem{15}Schneider, P., 1996, MNRAS, 283,837 
\bibitem{16}Trentham, N., M\"oller, O., Ramirez-Ruiz, E., 2001, MNRAS, 332,
658  
\bibitem{17}Umetsu, E. \& Futamase, T., 2000, ApJL, 539, L5
\bibitem{18}Vikhlinin, A., McNamara, B.R., Hornstrup, A. et al., 1999, ApJL,
520, L1
\bibitem{19}White, M., Van Waerbeke, L., Mackey, J., 2001, ApJ
submitted, astro-ph/0111490 
\bibitem{20}Wittman, D., Tyson, J.A., Margoniner, V.E., Cohen, J.G.,
Dell'Antonio, I.P., 2001, ApJL, 557, L89 
\end{thebibliography}
\end{document}